\documentstyle[10pt,emulateapj]{article}

\lefthead{BAUMGARTE \& SHAPIRO}
\righthead{EVOLUTION OF ROTATING SUPERMASSIVE STARS TO THE ONSET OF COLLAPSE}
\begin{document}

\title{EVOLUTION OF ROTATING SUPERMASSIVE STARS TO THE ONSET OF COLLAPSE}
\author{Thomas W.~Baumgarte and 
        Stuart L.~Shapiro\altaffilmark{1}}
\affil{Department of Physics, University of Illinois at
        Urbana-Champaign, Urbana, Il~61801}
\altaffiltext{1}{Department of Astronomy and
        National Center for Supercomputing Applications, 
        University of Illinois at Urbana-Champaign, Urbana, Il~61801}

\begin{abstract}
We launch a fully relativistic study of the formation of supermassive
black holes via the collapse of supermassive stars.  Here we initiate
our investigation by analyzing the secular evolution of supermassive
stars up to the onset of dynamical instability and collapse.  We focus
on the effects of rotation, assumed uniform, and general relativity.
We identify the critical configuration at which radial instability
sets in and determine its structure in detail.  We show that the key
nondimensional ratios $R/M$, $T/|W|$ and $J/M^2$ ($T$ is the
rotational kinetic energy and $W$ is the gravitational potential
energy) for this critical configuration are universal numbers,
independent of the mass, spin, radius or history of the star.  We
compare results from an approximate, analytic treatment with a fully
relativistic, numerical calculation and find good agreement. We solve
analytically for the time evolution of these parameters up to the
onset of instability.  Cooling by photon radiation drives the
evolution, which is accompanied by mass, angular momentum and entropy
loss.  The critical configuration serves as initial data for a future
relativistic, hydrodynamical, 3D simulation of the collapse of an
unstable supermassive star.  Since this implosion starts from a
universal critical configuration, the collapse is also uniquely
determined and should produce a universal gravitational waveform.  In
this paper we briefly speculate on the possible outcome of this
collapse and asses to what extent it offers a promising route to
forming a supermassive black hole.
\end{abstract}

\section{INTRODUCTION}

Recent observations provide increasingly strong evidence that
supermassive black holes (SMBHs) exist and that they are the sources
that power active galactic nuclei (AGNs) and quasars (see, e.g., Rees
1998, for a summary and references).  For example, the 1.3 cm
maser-emission line of H$_2$O allows for an extremely accurate mapping
of gas motions in a disk at the core of the galaxy NGC 4258.  The
inner edge of the disk has a radius of $\sim 0.1$ pc, and the inferred
velocity profile is consistent with Keplerian motion around a central
object of $3.6 \times 10^7 M_{\odot}$ (Claussen \& Lo 1986; Watson
\& Wallin 1994; Miyoshi et.~al. 1995).  The most conservative
candidate for so massive an object in so small a volume is a black
hole.

However, the scenario by which SMBHs form with masses in the range of
$10^6$--$10^{10} M_{\odot}$ is still uncertain (see Rees 1984 for an
overview).  Viable stellar dynamical and a hydrodynamical routes
leading to the formation of SMBHs have both been proposed.  In one
stellar dynamical scenario, the primordial gas first fragments into
stars, which form a dense stellar cluster or galaxy core.  Repeated
collisions and mergers of the stars in the cluster lead to the
build-up of massive stars, and the subsequent collapse of the massive
stars result in the formation of stellar-mass black holes. As these
holes continue to swallow stars and grow, they settle toward the
galaxy center, merge and trigger the build-up of one or more SMBHs
(see, e.g., Quinlan \& Shapiro 1990 and references therein).  In
another dynamical scenario, the gravothermal catastrophe (secular core
collapse) in a dense cluster composed of neutron stars or stellar-mass
black holes may drive the cluster core to a relativistic state.
Sufficiently relativistic clusters are dynamically unstable to
catastrophic collapse and may give birth to SMBHs (Zel'dovich \&
Podurets 1965; Shapiro \& Teukolsky 1985a, 1985b; Quinlan \& Shapiro
1987).  It may be more likely that the contracting primordial gas
builds up sufficient radiation pressure to inhibit further
fragmentation and hence prevent the formation of a stellar cluster.
Alternatively, a supermassive gas cloud may build up from the
fragmentation of stars in collisions in stellar clusters (Sanders
1970; Begelman \& Rees 1978).  In such hydrodynamical scenarios, a
massive black hole may form directly from gas, perhaps after a phase
in which the gas evolves in a quasi\-stationary manner as a
supermassive star (SMS).  Such a phase can only be transient, since
ultimately general relativity will drive such stars dynami\-cally
unstable (Iben 1963; Chandrasekhar 1964a, b; Feynman, unpublished, as
quoted in Fowler 1964).  The structure and evolution of supermassive
stars up to the onset of instability has been explored previously by
many authors, including Hoyle \& Fowler (1963a, b), Fowler (1964,
1966), Bisnovatyi-Kogan, Zel'dovich \& No\-vi\-kov (1967), Appenzeller
\& Fricke (1972) and Fuller, Woosley and Weaver (1986) (see Zel'dovich
\& No\-vi\-kov, 1971, and Shapiro and Teukolsky, 1983, for reviews and
further references).

In addition to their importance for a fundamental understanding of
AGNs and quasars, supermassive objects and the formation of SMBHs have
recently attracted new interest because they are likely candidates for
detection by currently proposed space-based gravitational wave
detectors like the Laser Interferometer Space Antenna (LISA).  Because
of its long arm length, this detector would be very sensitive to long
wavelength and low frequency radiation, and therefore supermassive
objects are among the most promising sources (see, e.g., Thorne \&
Braginsky 1976; Thorne 1995).  In particular, LISA may be able to
detect the collapse of a SMS to a SMBH.  Even more promising is the
possible detection of the coalescence of two SMBHs (LISA Pre-Phase A
report 1995).  The likelihood of such an event, however, largely
depends on how SMBHs form and is therefore still uncertain.

In a series of papers we revisit the formation of SMBHs via the
collapse of SMSs, focussing on the influence of rotation and general
relativity.  We analyze the secular contraction of a uniformly
rotating equilibrium configuration via thermal emission and mass
loss. We concentrate on a configuration rotating at the mass-shedding
limit.  In Baumgarte \& Shapiro (1999, hereafter Paper I), we have
shown that the luminosity from such a star is considerably reduced
below the value of a nonrotating spherical star of the same mass.  

In this paper, we analyze the structure and stability against collapse
of fully relativistic, rotating $n=3$ polytropes in stationary
equilibrium.  SMSs to which these calculations apply are
radiation-dominated, isentropic configurations of sufficient mass that
neither nuclear burning nor electron-positron pairs are important
before the stars reaches the onset of relativistic gravitational
instability.  Stars with $M \gtrsim 10^6 M_{\odot}$ fall in this
category (Zel'dovich \& No\-vi\-kov 1971; Fuller, Woosley \& Weaver 1986).
Moreover, the evolutionary timescale due to cooling has to be longer
than the hydrodynamic timescale for the star to evolve in a
quasistationary fashion.  According to equations (9) and (10) below,
we find that this constraint is satisfied for all masses
$M \lesssim 10^{13} M_{\odot}$.

We track the quasistationary evolution of such stars to the point of
onset of radial instability.  We first identify this critical
configuration analytically by means of a post-Newtonian energy
variational method.  As pointed out by Bisnovatyi-Kogan, Zel'do\-vich \&
No\-vi\-kov (1967), determining the onset of relativistic collapse in a
rapidly rotating SMS requires a {\em second-order} post-Newtonian
analysis, not a first-order, as in the case of a nonrotating star.  We
provide an approximate, post-Newtonian analytic analysis (rigorizing and
completing the qualitative argument of Bisnovatyi-Kogan, Zel'do\-vich \&
No\-vi\-kov (1967) and Zel'dovich \& No\-vi\-kov (1971), who only provide a
dimensional estimate of the second post-Newtonian term).  Then, to
establish the result rigorously, we use a recent numerical code (Cook,
Shapiro \& Teukolsky 1992, 1994) to construct rotating equilibrium
models in full general relativity.  We show that the mass of the
critical configuration is a unique function of the specific entropy.
The values of $R/M$, $T/|W|$ and $J/M^2$ at the onset of collapse are
universal numbers, independent of the mass or prior evolution.  Here
$M$ is the mass, $R$ the (polar) radius, $J$ the angular momentum, $T$
the rotational and $W$ the gravitational potential energy of the star.
We also speculate on the likely outcome of collapse for stars which do
not disrupt due to thermonuclear explosions during collapse; it is
found that stars with $M > 10^5 M_{\odot}$ and initial metallicities
$Z < 0.005$ do not explode (Fuller, Woosley \& Weaver 1986).  Since
these stars start collapsing from a universal critical
configuration, the subsequent collapse is also uniquely determined and
should produce a unique gravitational waveform.  We
postpone a detailed discussion of this phase for a future paper in
which we follow the dynamical collapse numerically in general
relativity (Baumgarte, Shapiro \& Shibata 1999).

The key goals of our study are to decide whether a SMBH can really
emerge from the collapse of a SMS and to determine the hole parameters
if indeed it can be formed this way.  Alternatively, a rotating
supermassive cloud or star could collapse to a weakly
relativistic disk (e.g.~Wagoner 1969; Loeb \& Rasio 1994).  If the
collapsing innermost region enters the strong-field domain, the
angular momentum of this matter must be below the maximum value of a
Kerr hole ($J/M^2 = 1$) for black hole formation to occur eventually.
What happens if the angular momentum exceeds this limit?  Does
angular momentum dissipation by outflowing gas allow for black hole
formation of the core? Or, does gravitational radiation, following the
formation of bars or axial currents carry away enough angular momentum
to permit collapse?  Finally, if a SMBH can form by the collapse of a
SMS, what is its mass and spin given the mass and spin of the SMS at
the onset of collapse?

In this paper we deal primarily with the structure, stability and
early secular evolution phases of the SMS scenario. Our calculation,
in effect, sets up the initial data at the onset of collapse.
Tracking the subsequent dynamical evolution of these initial data will
resolve the key questions posed above, and in this paper we will only
speculate briefly on the outcome of the dynamical collapse.

This paper is organized as follows: In Section~\ref{Sec2} we provide
a qualitative overview of the problem and present our basic
assumptions.  In Section~\ref{Sec3} we discuss the equilibrium and
stability of rotating, relativistic SMSs.  In particular, we determine
the critical configuration at which an evolving SMS becomes
dynamically unstable to radial perturbations.  We compare results from
an approximate analytical treatment (Section~\ref{anal}) with those
from a numerical, fully relativistic calculation
(Section~\ref{numerics}).  Having identified the onset of instability,
we then solve analytically for the evolution of SMSs during the
secular contraction phase up to this critical configuration in
Section~\ref{Sec4}.  In Section~\ref{coll} we provide some qualitative
arguments which suggest that the direct formation of SMBHs from the
collapse of SMSs indeed may be possible.  We summarize and discuss our
results in Section~\ref{Summary}.  Except where noted otherwise, we
adopt geometrized units with $c \equiv 1 \equiv G$ throughout this
paper.

\section{QUALITATIVE OVERVIEW AND BASIC ASSUMPTIONS}
\label{Sec2}

SMSs may form if collapsing primordial gas builds up enough entropy so
that the radiation pressure can slow down the collapse (see Begelman
\& Rees, 1978, for an alternative scenario).  Further contraction will
then spin up the newly formed SMS to the mass-shedding limit, provided
that the gas had some initial angular momentum and that viscosity
maintains uniform rotation.  The SMS will then evolve secularly along
the mass-shedding limit, simultaneously emitting radiation, matter and
angular momentum (see, e.g., Bisnovatyi-Kogan, Zel'dovich \& No\-vi\-kov
1967; Zel'dovich \& No\-vi\-kov 1971).  Once it reaches the onset of
radial instability, the star collapses on a dynamical timescale, and
may ultimately form a SMBH.

For sufficiently massive objects ($M \gtrsim 10^6 M_{\odot}$), the
equation of state is dominated by thermal radiation pressure.  It can
also be shown that SMSs are convective (see Loeb \& Rasio 1994, for a
simple proof) with constant entropy per baryon,
\begin{equation}
s \approx \frac{4}{3} \, \frac{a T^4}{n_{b}},
\end{equation}
where $n_{b}$ is the baryon density and $a$ is the radiation density
constant.  These conditions imply that the structure of a SMS is that
of an $n=3$ polytrope
\begin{equation}
P = K \rho^{4/3},
\end{equation}
where
\begin{equation}
K = K(s) = \frac{a}{3} \left(\frac{3 s}{4 m_H a} \right)^{4/3} =
	const.
\end{equation}
(see eq.~17.2.6 in Shapiro \& Teukolsky 1983). Here, $m_H$ is the mass
of a hydrogen atom and $K$ has been evaluated for a composition of pure 
ionized hydrogen.  In Newtonian theory, the mass of a static,
equilibrium $n=3$ polytrope is {\em uniquely} determined by the
polytropic constant $K$ alone
\begin{equation} \label{M_Newton_K}
M = \left( K \frac{k_1}{k_2} \right)^{3/2}
\end{equation}
(the numerical coefficients $k_1$ and $k_2$, derived from Lane-Emden
functions, are given in Table~1 below).  Inverting this relation, we
can write $K$ in terms of $M$ as
\begin{equation} \label{K}
K = 1.01 \times 10^3 \,\mbox{cm}^{2/3}
        \left( \frac{M}{M_{\odot}} \right)^{2/3},
\end{equation}
Since $n=3$ polytropes are extremely centrally condensed, the
self-gravity of the outermost envelope of a uniformly rotating
configuration can be neglected and the Roche approximation (see
Section~\ref{roche}) can be applied.

In Newtonian gravitation, $n=3$ polytropes are mar\-gin\-ally stable
to radial collapse.  However, even very small general relativistic
corrections make these polytropes unstable, and some mechanism has to
be invoked to prevent collapse.  In this paper, we focus on rotation,
which up to a critical configuration along the mass-shedding sequence
can stabilize SMSs.  In Section~\ref{Sec3} we will identify this
critical configuration (which we will call ``configuration $A$'').
Alternatively, gas pressure may stabilize SMSs in the absence of
angular momentum (see Zel'dovich \& No\-vi\-kov 1971; Shapiro \& Teukolsky
1983), but even a small degree of rotation will dominate gas pressure
(cf.~Section~\ref{anal}).  A dark matter background also tends to have
a stabilizing effect (Fuller, Woosley \& Weaver, 1986), but it is also
much less important than rotation (Bisnovatyi-Kogan 1998).  Since the
ratio between the rotational kinetic and the potential energy,
$T/|W|$, is always very small for a uniformly rotating star during
secular evolution along the mass-shedding sequence (see
eq.~[\ref{t_ms}] and Figure~\ref{fig4}), it is unlikely that nonradial
modes of instability are important during this phase.  Once the star
has become unstable and collapses, however, $T/|W|$ can become very
large, and it is likely that such nonradial modes develop (see
Section~\ref{coll}).

Massive enough stars ($M \gtrsim 10^6 M_{\odot}$) do not
reach sufficiently high temperatures for nuclear burning to become
important before reaching the onset of instability
(see~Section~\ref{Sec3.4}).  Also, electron-positron pairs play a
negligible role in this regime (Zel'dovich \& No\-vi\-kov 1971).  The
evolution of the SMS along the mass-shedding sequence is then
determined solely by cooling via thermal photon emission and loss of
mass and angular momentum.  As we have shown in Paper I, the
luminosity of a rotating supermassive star at mass-shedding is
\begin{equation}
L = 0.639\, L_{\rm Edd},
\end{equation}
where the Eddington luminosity is
\begin{equation} \label{eddington}
L_{\rm Edd} = \frac{4 \pi M}{\kappa}.
\end{equation}
The opacity $\kappa$ is dominated by Thomson scattering off free
electrons
\begin{equation}
\kappa = \kappa_{T} = 0.2\, (1 + X_H) \,\mbox{cm}^2\mbox{g}^{-1},
\end{equation}
where $X_H$ is the hydrogen mass fraction.  For $M \gtrsim 10^5
M_{\odot}$, temperatures are low enough that Klein-Nishina
corrections can be neglected (see, e.g., Fuller, Woosley \& Weaver
1986).

As we will find in Section~\ref{Sec4}, the evolutionary timescale
for secular evolution along the mass-shedding sequence is given
by
\begin{equation}
t_{\rm evol} = 8.8 \times 10^{11} \mbox{s} \left( \frac{R/M}{456} \right)^{-1}
\end{equation}
where the coefficient has been evaluated near the critical
configuration (see eq.~(\ref{tcrit_value})).  The dynamical
timescale is
\begin{eqnarray}
t_{\rm dyn} & \sim & (G \rho_c)^{-1/2} \\
& = & 1.3 \times 10^{-2} \mbox{s}
	\left( \frac{M}{M_{\odot}} \right)
        \left( \frac{\bar \rho_c}{ 7 \times 10^{-9}} \right)^{-1/2},\nonumber
\end{eqnarray}
where we have used eq.~(\ref{rho_value}) below to evaluate $\rho_c$
in terms of the non-dimensional density $\bar \rho_c$.
Given the critical values for $\bar \rho_c$ and $R/M$ that we find in
Section~\ref{Sec3}, the evolutionary timescale is longer than the
dynamical timescale for $M \lesssim 6.7 \times 10^{13}
M_{\odot}$. This mass is larger than the value derived in Shapiro \&
Teukolsky (1983), where a similar estimate has been made for
nonrotating SMSs stabilized by gas pressure.

Also important is the viscous timescale, which is very uncertain.  Not
surprisingly, the microscopic viscosity due either to collisions
between ions or to radiation yields timescales which are larger than
the evolutionary timescale by many orders of magnitude (see Kippenhahn
\& Weigert 1990).  The effect of turbulent viscosity can be estimated
by assuming that the velocity of the turbulent motion $v_t$ is an
appreciable fraction of the velocity of sound,
\begin{equation}
v_t = \alpha v_{\rm sound},
\end{equation}
where we take the dimensionless viscosity parameter $\alpha$ to lie in
the range
\begin{equation}
0.01 \lesssim \alpha \lesssim 1
\end{equation}
(see Shakura \& Sunyaev 1973; also Zel'dovich \& No\-vi\-kov 1971;
Balbus \& Hawley 1991).  Assuming the characteristic scale of
nonuniformity to be some fraction of the stellar radius $R$, we can
then estimate the turbulent viscosity to be
\begin{equation}
\eta_{t} \sim \rho R v_t = \rho R \alpha v_{\rm sound}.
\end{equation}
The corresponding timescale is then
\begin{equation}
t_{\rm vis} \sim \frac{R^2 \rho}{\eta_t} \sim \frac{R}{\alpha v_{\rm sound}}
\sim \alpha^{-1} t_{\rm dyn}.
\end{equation}
It has been argued that both convection and magnetic fields serve to
generate such a turbulent viscosity in SMSs (Bisnovatyi-Kogan,
Zel'dovich \& No\-vi\-kov 1967; Wagoner 1969).

For SMSs with masses $M \ll 6.7 \times 10^{11} M_{\odot}$, we 
therefore find the hierarchy 
\begin{equation} \label{time_hier}
t_{\rm dyn} \ll t_{\rm vis} \ll t_{\rm evol}.
\end{equation}
Eq.~(\ref{time_hier}) justifies our assumption that the star evolves
in a quasistationary manner along the mass-shedding sequence.  The
last inequality implies that turbulent viscosity, if present, will
keep the SMS in uniform rotation. We will assume this result in
treating the secular evolution up to the onset of instability.  Once
the star becomes unstable and starts collapsing, however, the
evolution will proceed on the dynamical timescale, and it is unlikely
that the viscosity will be efficient enough to maintain uniform
rotation.  It is more likely that cylindrical mass shells
will conserve their angular momentum, which we assume for our
qualitative arguments in Section~\ref{coll}.

\section{EQUILIBRIUM AND STABILITY}
\label{Sec3}

In this Section we analyze the equilibrium and stability of rotating
SMSs and identify the critical configuration at which radial
instability sets in. We briefly review the predictions of the Roche
approximation in Section~\ref{roche}.  In Section~\ref{anal} we
present an analytic, post-Newtonian model calculation, which provides
qualitative insight into the scaling behavior and approximate numerical
parameters for the critical configuration.  In Section~\ref{numerics},
we present numerical results from a fully relativistic calculation,
which allows us to identify the critical configuration and its
characteristic parameters (including $J/M^2$ and $R/M$) independently of
the approximations used in the analytical treatment (including the
post-Newtonian and Roche approximation).  We find remarkable agreement
between the two calculations.  Lastly, in Section~\ref{Sec3.4}, 
we restore cgs units and provide expressions for the physical
parameters in SMSs. 

\subsection{Predictions of the Roche Model}
\label{roche}

The Roche approximation for rotating $n=3$ polytropes has been derived
and discussed in the literature (see, e.g., Zel'dovich \&
No\-vi\-kov 1971; Shapiro \& Teukolsky 1983). Here we briefly summarize
the key elements for application below (see also Paper I).

Stars with soft equations of state are extremely centrally condensed:
they have an extended, low density envelope, with the bulk of the mass
concentrated in the core.  For an $n=3$ polytrope, for example, the
ratio between central density to average density is $\rho_c/ \langle
\rho
\rangle = 54.2$.  The gravitational forces in the envelope are
therefore dominated by the massive core, and it is appropriate to
neglect the self-gravity of the envelope.  In this case, it is easy to
show that matter at the equator of a rotating star orbits with the Kepler
frequency when the ratio between the equatorial radius $R_e$ and the
polar radius $R_p$ reaches
\begin{equation}
\left( \frac{R_e}{R_p} \right)_{\rm shedd} = \frac{3}{2}.
\end{equation}
At this point, the angular velocity of the star assumes its maximum
value before mass-shedding from the equator sets in:
\begin{equation} \label{Omega_shedd}
\Omega_{\rm shedd} =
\left( \frac{2}{3} \right)^{3/2} \left( \frac{M}{R_p^3} \right)^{1/2}.
\end{equation}
The polar
radius hardly changes from the value of the radius of the
nonrotating, spherical star, a result which has been verified numerically
(Papaloizou \& Whelan 1973).  Since the bulk of the matter is
concentrated at the core and hardly affected by the rotation,
the moment of inertia of inertia of the star barely changes with
rotation and is well approximated by the nonrotating
value
\begin{equation} \label{I}
I = \frac{2}{3} \langle r^2 \rangle M
	= \frac{2}{3} \,0.1130 \,R_p^2 M,
\end{equation}
where the brackets denote a mass weighted average over the star, and
where the last equality holds for $n = 3$ (this approximation turns
out to be less accurate; cf.~Section~\ref{numerics}).  The ratio
between the kinetic and potential energy at mass-shedding then becomes
\begin{eqnarray} \label{t_ms}
\left( \frac{T}{|W|} \right)_{\rm shedd} & = &
	\frac{(1/2)I\,\Omega_{\rm shedd}^2}{(3/2)M^2/R_p} \nonumber \\
	& = & \left( \frac{2}{3} \right)^4 \frac{0.1130}{3} = 0.00744.
\end{eqnarray}
This result predicts that $T/|W|$ of a maximally rotating, $n = 3$
polytrope (or any star for which the Roche approximation applies) is a
universal constant, independent of mass, radius, or orbital velocity.
In Section~\ref{numerics} we will see that precise numerical models
predict a somewhat larger value of $T/|W|$, which nevertheless also
remains approximately constant along a mass-shedding trajectory
(see Figure~\ref{fig4} below).

\subsection{Analytical Model}
\label{anal}

To determine the equilibrium and stability of a rotating SMS, we write
its total energy as the sum of the internal
energy $U$, the potential energy $W$, the rotational energy $T$, a
post-Newtonian correction $E_{PN}$ and a post-post-Newtonian
correction $E_{PPN}$.  For a $n=3$ polytrope, these terms can be
written
\begin{eqnarray} \label{energy}
E & = & k_1 K M \rho_c^{1/3} - k_2 M^{5/3} \rho_c^{1/3} + \\[1mm]
	& & k_3 j^2 M^{7/3} \rho_c^{2/3} 
	- k_4 M^{7/3} \rho_c^{2/3} - k_5 M^3 \rho_c, \nonumber
\end{eqnarray}
where we have defined $j \equiv J/M^2$ and have neglected corrections
due to deviations from sphericity.  This neglect is justified, since
these corrections scale with $T/|W|$, which according to~(\ref{t_ms})
is always very small.  Even though the value of the
post-post-Newtonian correction $E_{PPN}$ is very small, this term is
crucial for determining the critical, marginally stable configuration,
as emphasized by Zel'dovich and No\-vi\-kov (1971).  The values of the
nondimensional coefficients $k_i$ are listed in Table~1; they are
constructed from Lane-Emden functions.

\begin{deluxetable}{cccccc}
\tablewidth{0pc}
\tablecaption{Values of the structure coefficients for $n=3$.}
\tablehead{ $k_1$ & $k_2$ & $k_3$ & $k_4$ & $k_5$ } 
\startdata 
1.7558\tablenotemark{1}	& 0.63899\tablenotemark{1}
	&  1.2041\tablenotemark{1} & 0.918294\tablenotemark{2} &
	0.331211\tablenotemark{3}
\tablerefs{(1) Lai, Rasio \& Shapiro, 1993; (2) Shapiro \& Teukolsky, 1983;
	(3) Lombardi, 1997}
\enddata
\end{deluxetable}

Note that for any polytrope $K^{n/2}$ has units of length.  We can
therefore introduce nondimensional coordinates by setting $K = 1$ (see
also Cook, Shapiro \& Teukolsky 1992).  We will denote values
of nondimensional variables in these coordinates with a bar
(for example $\bar M$).  Values of these quantities for any other 
value of $K$ can be recovered easily by rescaling with an appropriate
power of $K^{n/2} = K^{3/2}$; for example $M = K^{3/2} \bar M$ and
$\rho = K^{-3} \bar \rho$.

Taking the first derivative of~(\ref{energy}) with respect to the
central density yields a condition for equilibrium
\begin{eqnarray} \label{equilibrium}
0 = \frac{\partial \bar E}{\partial x} & = & 
	k_1 \bar M - k_2 \bar M^{5/3} + 2 k_3 j^2 \bar M^{7/3} x - 
	\nonumber \\[1mm]
	& &	2 k_4 \bar M^{7/3} x - 3 k_5 \bar M^3 x^2,
\end{eqnarray}
where $x = \bar \rho_c^{1/3}$.  For stable equilibrium, the second
derivative of eq.~(\ref{energy}) has to be positive.  A root of the second
derivative therefore marks the onset of radial instability
\begin{equation} \label{stability}
0 = \frac{\partial^2 \bar E}{\partial x^2} =
	2 k_3 j^2 \bar M^{7/3} 
	- 2 k_4 \bar M^{7/3} - 6 k_5 \bar M^3 x.
\end{equation} 
Solving eq.~(\ref{stability}) for $x$ immediately yields
\begin{equation} \label{x}
\bar \rho_c^{1/3} = x = \frac{k_3 j^2 - k_4}{3 k_5 \bar M^{2/3}}.
\end{equation}
For stability, we obviously require $\bar \rho_c \geq 0$ or
\begin{equation}
j \geq j_{\rm min} = \left( \frac{k_4}{k_3} \right)^{1/2} = 0.8733.
\end{equation}
Note that stable configurations with $j = j_{\rm min}$ have $\bar
\rho_c = 0$. 

SMSs with angular momenta less than $j_{\rm min}$ can also be
stabilized by the nonvanishing contribution of the plasma, which
effectively decreases the polytropic index $n$ to a value slightly
smaller than 3 (see Shapiro \& Teukolsky 1983, Chapt.~17).  Here,
because we are interested in the effects of rotation, we neglect the
plasma and approximate supermassive stars as strict $n=3$ radiation
dominated polytropes.  Our neglect of gas pressure breaks down for angular
momenta less than $j_{\rm min}$, since it would make wrong predictions
for the stability of supermassive stars.  We expect that most
equilibria which form in nature have $j \gg j_{\rm min}$.

Inserting~(\ref{x}) into~(\ref{equilibrium}), we find the mass of 
equilibrium stars at the onset of instability is
\begin{equation} \label{M}
\bar M^{2/3} = k_1 \left[ k_2 - \frac{1}{3k_5} 
	\left( k_3 j^2 - k_4 \right)^2 \right]^{-1}.
\end{equation}
For $j = j_{\rm min}$, and hence $\bar \rho_c = 0$, the configuration
has infinite radius and we obviously
recover the (unique) Newtonian mass of a $n=3$ polytrope
\begin{equation} \label{M_Newton}
\bar M_{\rm min} = \left( \frac{k_1}{k_2} \right)^{3/2} = 4.5548
\end{equation}
(where we keep enough digits for purposes of comparison later on).
This configuration corresponds to point $B$ in Figs.~\ref{fig2}
and~\ref{fig3}.  Note that from equation~(\ref{M}), all critical
configurations at the onset of instability have a mass greater or
equal to $\bar M_{\rm min}$.

The maximum angular velocity of a rotating star occurs at
mass-shedding, where we know the value of $T/|W|$ (eq.~\ref{t_ms} in
the Roche approximation). It is convenient to express $T/|W|$ in
terms of $x$, $\bar M$ and $j$
\begin{equation} \label{t}
\frac{T}{|W|} = \frac{k_3 j^2 \bar M^{7/3} x^2}{ k_2 \bar M^{5/3} x}
	= \frac{k_3 j^2 \bar M^{2/3} x}{ k_2 }.
\end{equation}
The last stable configuration for a given $T/|W|$ can now be found by
substituting~(\ref{x}) into~(\ref{t}), which yields a quadratic 
equation for $j^2$:
\begin{equation}
j^4 - \frac{k_4}{k_3} j^2 - 3 \frac{k_2 k_5}{k_3^2} \frac{T}{|W|} = 0,
\end{equation}
or
\begin{equation}
j^2 = \frac{k_4}{2 k_3} + \sqrt{\frac{k_4^2}{4 k_3^2} +
	3 \frac{k_2 k_5}{k_3^2}\frac{T}{|W|} }.
\end{equation}
(The second solution can be disregarded, since it gives a negative
value for $j^2$).  To first order in $T/|W|$, this result can be
approximated by
\begin{equation} \label{j_crit}
j = j_{\rm min} (1 + \frac{3}{2} \frac{k_2 k_5}{k_4^2}\,\frac{T}{|W|} )
	= j_{\rm min} (1 + 0.377 \,\frac{T}{|W|}).
\end{equation}
The mass of this critical configuration now can be found by
inserting~(\ref{j_crit}) into~(\ref{M}), which yields
\begin{equation}
\bar M^{2/3} = \frac{k_1}{k_2} 
\left(1 - 3 \frac{k_5 k_2}{k_4^2} \Big(\frac{T}{|W|}\Big)^2 \right)^{-1},
\end{equation}
or, to first order in $(T/|W|)^2$,
\begin{eqnarray} \label{M_crit}
\bar M & = & \bar M_{\rm min} \left( 1 + \frac{9}{2} \frac{k_5 k_2}{k_4^2} 
	\Big(\frac{T}{|W|}\Big)^2\right) \nonumber \\
 	& = &  \bar M_{\rm min} \left(1 + 1.1294 \,\Big(\frac{T}{|W|}\Big)^2
	\right).
\end{eqnarray}
Adopting the Roche prediction~(\ref{t_ms}) for $T/|W|$, we find
\begin{eqnarray} \label{crit_values}
j_{\rm crit} & = & 0.8757 \label{j_crit_num} \\
\bar M_{\rm crit} & = & 4.5551.
\end{eqnarray}
Inserting these values into~(\ref{x}) yields
\begin{equation}
(\bar \rho_{c})_{\rm crit} = 6.6 \times 10^{-9}.
\end{equation}
These values correspond to the critical configuration $A$ in
Figs.~\ref{fig2} and~\ref{fig3}.

\begin{deluxetable}{llllllllll}
\tablewidth{0pc}
\tablecaption{Critical values at point $A$.}
\tablehead{ & $\bar \rho_c$ & $\bar M$ & \colhead{$R_p/M$} 
	& $J/M^2$ & $T/|W|$ 
	& $R_p/R_e$ & $\Omega/\Omega_{\rm spher}$\tablenotemark{a}
	& $I/I_{\rm spher}$\tablenotemark{b} }
\startdata 
Analytical Model	& $6.6 \times 10^{-9}$	
	& 4.5551 & 456 	& 0.8757 & $7.44 \times 10^{-3}$ 
	& 0.6667  & 0.5443 & 1 \nl
Numerical Value		& $7 \times 10^{-9}$	
	& 4.57	& 427	& 0.97 	 & $8.99 \times 10^{-3}$ 
	& 0.664  & 0.5441 & 1.15
\tablenotetext{a}{$\Omega_{\rm spher} \equiv (M/R_p^3)^{1/2}$}
\tablenotetext{b}{$I_{\rm spher} \equiv \frac{2}{3}\,0.1130\,R_p^2 M$}
\enddata
\end{deluxetable}

It is obvious from expressions~(\ref{j_crit}) and~(\ref{M_crit}) that
the critical values $j_{\rm crit}$ and $\bar M_{\rm crit}$ depend on
several approximations.  In this section, we have adopted the Roche
prediction~(\ref{t_ms}) for $T/|W|$ at mass-shedding, even though
self-consistent numerical calculations yield a value that differs by
about 20\% (cf.~Section~\ref{numerics} and Table~2).  Moreover, the
critical values depend on the second post-Newtonian coefficient $k_5$.
We have adopted a value that only takes into account second
post-Newtonian corrections to the energy of a spherical, nonrotating
configuration (Lombardi 1997), and have neglected first
post-Newtonian corrections to the Newtonian rotational kinetic
energy. The later would have the same form as the second
post-Newtonian term $E_{PPN}$, and would therefore introduce an
unknown correction to the coefficient $k_5$ (which we expect to be
small for small values of $T/|W|$).  Together, this means that our
analytic calculation can only yield approximate values for the
critical configuration.

It is interesting to note, however, that our analytic model predicts 
$j_{\rm crit}$ to be less than unity, which is the maximum value allowed
for a Kerr black hole.  A supermassive star that 
evolves along the mass-shedding sequence and ultimately becomes
unstable could therefore collapse to a black hole without having to 
loose additional angular momentum (see also Section~\ref{coll}).

Lastly, we estimate the compaction of the critical configuration by
rewriting equation~(\ref{t}) in terms of the polar radius $\bar R_p$.
Neglecting deviations from sphericity, we find
\begin{equation} \label{t2}
\frac{T}{|W|} = \frac{2 \bar J^2 \bar R_p}{6 \bar I \bar M^2}
	= \frac{1}{2 \times 0.1130} \frac{\bar M}{\bar R_p} j^2
\end{equation}
or
\begin{equation}
\frac{R_p}{M} =  4.425 \frac{j^2}{T/|W|}, 
\end{equation}
where we have dropped the bars on the left hand side, since $R_p/M$ is
a dimensionless combination, and have used eq.~(\ref{I}) to evaluate
$I$.  Inserting~(\ref{t_ms}) 
and~(\ref{j_crit_num}), we find
\begin{equation} \label{romcrit}
\left( \frac{R_p}{M} \right)_{\rm crit} = 456.
\end{equation}
This shows that the critical configuration is only very mildly relativistic, 
and that a post-Newtonian approximation is adequate for determining its
equilibrium structure.

The small admixture of thermal gas pressure can stabilize a
nonrotating SMS for values of $R/M$ exceeding $(R/M)^{\rm gas} = 1.59
\,(M/M_{\odot})^{1/2}$ (eq.~(17.4.11) in Shapiro \& Teukolsky, 1983).
For masses $M \gtrsim 10^5 M_{\odot}$, this ratio is larger than that
given eq.~(\ref{romcrit}).  Consequently, rotation serves to stabilize
these masses beyond the point up to which thermal pressure alone can
stabilize them.  Hence thermal gas pressure can be ignored, as we have
done, in determining the point of onset of instability in
mass-shedding configurations with $M \gtrsim 10^5 M_{\odot}$.

Note that all the nondimensional parameters of the critical configuration,
including $T/|W|$, $R/M$ and $J/M^2$, are universal constants, and do not
depend on the mass of the SMS or its history.

\subsection{Numerical Results}
\label{numerics}

\begin{figure*}
\epsscale{0.4}
\plotone{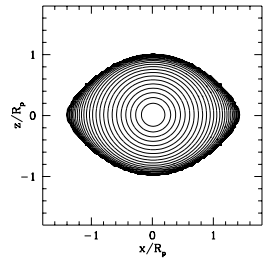}
\caption{Contours of constant mass energy density $\bar \rho$ in a plane
containing the axis of rotation for a $n=3$ polytrope at the
critical point $A$ (see Figs.~\protect\ref{fig1}
and~\protect\ref{fig2}). Each contour marks a change of the density by
a factor of $10^{1/3}$.\label{fig1}}
\end{figure*}

\begin{figure*}
\epsscale{0.4}
\plotone{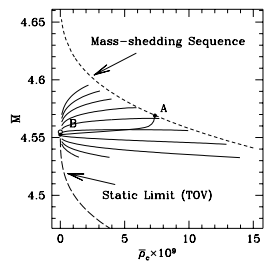}
\caption{Mass versus central density plot for relativistic, rotating
$n=3$ polytropes.  The long dashed curve is the TOV solution for nonrotating,
static configurations, and the short dashed curve marks the
mass-shedding limit.  The thin solid lines denote sequences of
constant angular momentum, ranging from $\bar J = 15$ (lowest curve)
to $\bar J = 24$ (highest curve) in increments of $\Delta \bar J = 1$.
Turning points of these curves mark the onset of instability. The
thick solid line connects these turning points (see also
Figure~\protect\ref{fig3}) and hence separates a region of stable
configurations (above this line) from a region of unstable
configurations (below this line).  In particular, all nonrotating
$n=3$ polytropes are unstable to radial perturbations.  A
configuration evolving along the mass-shedding sequence with
increasing central density becomes unstable at the critical point $A$.
All sequences of constant angular momentum connect the mass-shedding
limit with point $B$ for $\rho_c \rightarrow 0$ (and hence $R
\rightarrow \infty$).  The mass of this configuration should agree
with the mass $\bar M = 4.5525$ of a Newtonian $n=3$ polytrope
(equation~(\protect\ref{M_Newton})), which we have marked by the open
circle.  The deviation of the solid point $B$ from the analytical
value is a measure of our numerical accuracy. \label{fig2} }
\end{figure*}

\begin{figure*}
\epsscale{0.4}
\plotone{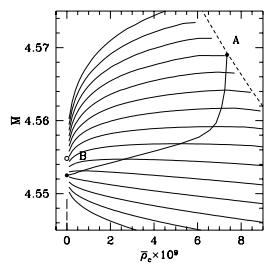}
\caption{Blowup of Figure~\protect\ref{fig2}.  Here, the thin solid
lines denote sequences of constant angular momentum raging from $\bar
 J = 17.5$ (lowest curve) to $\bar J = 21$ (highest curve) in
 increments of $\Delta \bar J = 0.25$. \label{fig3}}
\end{figure*}

In this Section we present numerical models of fully relativistic,
uniformly rotating $n=3$ polytropes.  We have constructed these models
with the numerical code of Cook, Shapiro \& Teukolsky (1992, 1994).
As a typical result, we show density contours of the last stable
(critical) configuration (configuration $A$) in a plane containing the
axis of rotation in Figure~\ref{fig1}.

In the static limit, we recover the Tolman-Op\-pen\-hei\-mer-Volkoff
result for relativistic hydrostatic equilibrium in spherical symmetry.
We mark this curve by the long dashed line in Figures~\ref{fig2}
and~\ref{fig3}.  Note that for a polytropic index $n=3$, Newtonian
gravity predicts a unique value for the mass $\bar M$, which is
independent of the central density (equation~(\ref{M_Newton})).  In
general relativity, $\bar M$ is no longer independent of $\bar
\rho_c$, but obviously we recover the the Newtonian value in the
Newtonian limit ($\bar \rho_c
\rightarrow 0$, point $B$ in Figures~\ref{fig2} and~\ref{fig3}).
Note that in general relativity all static $n=3$ polytropes are unstable
to radial perturbations ($\partial \bar M/\partial \bar \rho_c < 0$).

For each central density $\bar \rho_c$, there is a unique (maximum)
angular momentum, at which the equator reaches the Kepler frequency,
and at which mass-shedding sets in.  We mark this mass-shedding curve
by the short dashed line in Figures~\ref{fig2} and~\ref{fig3}. 

In order to find a transition from stable equilibrium to unstable
equilibrium, we need to construct sequences of constant angular
momentum and locate their turning points (Friedman, Ipser \& Sorkin,
1988).  We have marked sequences of constant angular momentum with
thin, solid lines in Figures~\ref{fig2} and~\ref{fig3}.  They all
connect the mass-shedding sequence with point $B$.  The later can be
understood by observing that, from equation~(\ref{t2}), $T/|W|$ scales
with $\bar R^{-1}$ for constant $j$.  As $\bar \rho_c \rightarrow 0$,
the mass $\bar M$ remains finite, and hence $\bar R \rightarrow
\infty$.  We therefore have $T/|W| \rightarrow 0$, so that rotation
plays an increasingly negligible role and we recover, for any $j$, the
static limit.

The sequences of constant angular momentum have a turning point only
for a small range of angular momenta ($18.25 \lesssim \bar J \lesssim
20$).  Above these values, the sequences are monotonically increasing,
while below they are monotonically decreasing.  The thick solid line
in Figures~\ref{fig2} and~\ref{fig3} connects the turning points, 
and therefore separates a region of stable equilibrium
(above the line), from a region of unstable equilibrium (below the line).

As it cools, a supermassive star evolves toward higher density along the
mass-shed\-ding sequence and is therefore stabilized by rotation
until it reaches point $A$.  At this point, the star becomes unstable
to radial perturbations and will collapse.  We summarize the numerical
parameters characterizing point $A$ in Table~2, where they are
compared with the findings of the analytical model calculation of
Section~\ref{anal}.  Figure~\ref{fig1} shows the density profile at
the critical point $A$ in a plane containing the axis of rotation.

\begin{figure*}
\epsscale{0.4}
\plotone{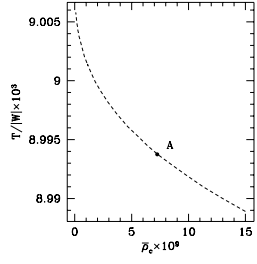}
\caption{$T/|W|$ as a function of central density 
along the mass-shedding line.  The dot marks the critical
configuration $A$ (see Figs.~\protect\ref{fig2}
and~\protect\ref{fig3}).  The numerical calculation yields a value of
$T/|W|$ which is different from the value $(T/|W|)_{\rm shedd} =
0.00744$ that the Roche approximation predicts.  Note, however, that
$T/|W|$ changes by far less than a percent over the region of interest,
and can therefore well be approximated to be constant. \label{fig4}}
\end{figure*}

Note that the numerical accuracy of an individual stellar model is
fairly high.  From the agreement of the limiting value $B$ (solid dot
in Figures~\ref{fig2} and~\ref{fig3}) with the theoretical value (open
circle), we estimate the numerical errors to be much less than 1 \%.
Identifying the critical point $A$, however, requires locating the
turning points along fairly flat curves, which therefore introduces a
larger error.  Nevertheless, we expect that we can identify the
characteristical parameters of point $A$ to, typically, within less
than 5\%.  Some of these parameters, however, change only very little
along the mass-shedding sequence and can therefore by determined to
much higher accuracy (for example $T/|W|$).  The number of digits in
Table~2 reflects the uncertainty in each parameter.

Having constructed self-consistent models of rotating $n=3$
polytropes, we can now evaluate the quality of the predictions of the
Roche approximation (see Section~\ref{roche}).  We find that the
predictions concerning the envelope alone, namely the ratio between
equatorial and polar radius and the orbital frequency at mass
shedding, are very accurate (Table 2).  Figure~\ref{fig1} also shows
that, at mass-shedding, only the envelope is distorted, while the core
of the star remains nearly spherical.  To predict $T/|W|$ at
mass-shedding, the Roche approximation assumes the moment of inertia
as well as the potential energy of the distorted star to be that of
the corresponding spherical star, which yields $T/|W| = 0.00744$,
independent of $\bar \rho_c$.  In reality, the moment of inertia will
be larger for the rotating star (by about 15\% for configuration $A$,
see the last column in Table~2), and the magnitude of the potential
energy slightly smaller.  Accordingly, we find that the Roche
approximation underestimates $T/|W|$ by about 20\%.  We also find that
$T/|W|$ is not independent of $\bar \rho_c$ (see Figure~\ref{fig4}).
However, over our region of interest, $T/|W|$ changes by much less than
1\%, so that it can very well be approximated by\footnote{Note that
the relativistic definitions used to calculate $T$ and $W$ numerically
are given in Cook, Shapiro \& Teukolsky (1992), eq.~(56) and~(65)}
$T/|W| = 0.009$ (cf.~Table 1 in Hachisu 1986).

\subsection{Numerical Values in cgs units}
\label{Sec3.4}

To obtain numerical values in cgs units from the dimensionless
``barred'' quantities, we have to restore the polytropic constant $K$
as well as $c$ and $G$.  Using eq.~(\ref{K}) for $K$, we find for the
density $\rho$
\begin{eqnarray} \label{rho_value}
\rho_c  & = & K^{-3} \bar \rho_c \\[1mm]
& = & 9 \times 10^{10} \mbox{\,g\,cm}^{-3}
	\left( \frac{M}{M_{\odot}} \right)^{-2} 
	\left( \frac{\bar \rho_c}{7.0 \times 10^{-9}} \right),
	\nonumber
\end{eqnarray}
where $7.0 \times 10^{-9}$ is the approximate value of the central
density of the critical configuration $A$.  The temperature can now be
found, for example, from eq.~(17.2.9) in Shapiro \& Teukolsky (1983)
\begin{equation}
T_c = 9 \times 10^{10} \mbox{K} \left( \frac{M}{M_{\odot}} \right)^{-1/2} 
	\left( \frac{\bar \rho_c}{7.0 \times 10^{-9}} \right)^{1/3}.
\end{equation}
For SMSs with masses $M \gtrsim 10^6 M_{\odot}$, the central
temperature is always less than $6 \times 10^7 K$.  According to Fowler
(1966), this is the minimum temperature required for generating the
SMS's luminosity via the CNO cycle.  This justifies our assumption
that nuclear reactions can be neglected.

\section{QUASISTATIONARY EVOLUTION TO THE ONSET OF INSTABILITY}
\label{Sec4}

In this Section we discuss the evolution of a rotating SMS up to the 
onset of instability.  In particular, we derive analytic expressions for
the evolution of the mass, radius and angular momentum as a function of 
time.  

The evolution of the three quantities is not independent. Instead,
they are coupled by two relations, namely the requirement that the
star evolves along mass-shedding (so that $T/|W|$ remains
approximately constant), and that the angular momentum loss can be
computed from the mass that leaves that star from the equator with the
critical angular velocity.  This means that we can express, for
instance, the angular momentum loss $\dot J$ and the change of radius
$\dot R$ in terms of the mass loss $\dot M$.  The only relation that
is yet to be determined fixes $\dot M$ and hence the overall timescale
for the evolution of the three quantities.

In a similar calculation, Bisnovatyi-Kogan, Zel'do\-vich \& No\-vi\-kov
(1967) estimated $\dot M$ by constructing an approximate stellar wind
model, which they joined onto the outer envelope of the star.  Their
model depends on several unknown nondimensional parameters dealing
with the wind solution.  Here, we take a much more naive approach,
determining the mass loss rate from the requirement that the star
remain in equilibrium as it cools in a quasistationary manner.
According to eq.~(\ref{M_Newton_K}), the mass loss rate is related to
the change of $K$ (or the entropy), and is thus governed by the star's
luminosity (see eq.~(\ref{mdot2}) below).

As we have seen in Section~\ref{anal}, the post-Newtonian corrections
and rotational contributions to the energy functional~(\ref{energy})
are important for determining the stability of SMSs, but they have a
very small effect on the equilibrium structure and can therefore be
neglected for the purposes of this Section.  Accordingly, the mass $M$
of the star is well approximated by the Newtonian
expression~(\ref{M_Newton_K}), which only depends on the polytropic
constant $K$.  The time derivative of the mass is given by
\begin{equation}
\frac{\dot M}{M} = \frac{3}{2} \, \frac{\dot K}{K}.
\end{equation}
The change of the total entropy $S$ of the star is related the
luminosity by
\begin{equation}
L = - T \dot S.
\end{equation}
Using the first law of thermodynamics, the right hand side can be rewritten
\begin{equation}
T \dot S = \int (T \dot s) dm = \dot K \int n \rho^{1/n} dm
	= U \, \frac{\dot K}{K},
\end{equation}
where $U$ is the internal energy (see Baumgarte \& Shapiro 1998,
eq.~[A38]).  For $n=3$ polytropes in equilibrium, the internal energy
is equal to the negative of the gravitational potential energy
\begin{equation}
U = - W = \frac{3}{2} \, \frac{M^2}{R}
\end{equation}
(see eq.~(\ref{equilibrium}), when post-Newtonian and rotational
corrections are neglected).  Collecting terms, we now find
\begin{equation} \label{mdot2}
\frac{\dot M}{M} = - \frac{L}{M} \, \frac{R}{M}.
\end{equation}

The luminosity $L$ can be written as
\begin{equation} \label{L1}
L = \lambda L_{\rm Edd},
\end{equation}
where the Eddington luminosity $L_{\rm Edd}$ has been defined 
in~(\ref{eddington}).
As we have shown in Paper I, the luminosity of a SMS, rotating at
the mass-shedding limit, is reduced by about 36\%
\begin{equation}
\lambda_{\rm shedd} = 0.639.
\end{equation}
Inserting~(\ref{L1}) into~(\ref{mdot2}), we now find
\begin{equation} \label{mdot3}
\frac{\dot M}{M} = - \frac{4 \pi \lambda}{\kappa} \, \frac{R}{M}.
\end{equation}

As we have seen in Section~\ref{Sec3}, a SMS that evolves along 
the mass-shedding limit conserves $T/|W|$ to very good
approximation:
\begin{equation}
\frac{T}{|W|} \sim const.
\end{equation}
Since $T/|W|$ is proportional to $J^2M^{-3}R^{-1}$, this implies
\begin{equation} \label{rdot1}
\frac{\dot R}{R} = 2\,\frac{\dot J}{J} -3\, \frac{\dot M}{M}.
\end{equation}

Assuming that the mass-shedding occurs at the equator, we can write
\begin{equation}
\dot J = l_{\rm esc} \dot M,
\end{equation}
where $l_{\rm esc}$ is the specific angular momentum of the escaping
matter
\begin{equation}
l_{\rm esc} = \Omega_{\rm shedd} R_e^2 = 
\frac{9}{4}\, \Omega_{\rm shedd} R_p^2.
\end{equation}
Since $J$ can be written
\begin{eqnarray}
J & = & I \Omega_{\rm shedd} = 
	\frac{2}{3} \,\langle r^2 \rangle \,M \Omega_{\rm shedd} \nonumber \\
& = & \frac{2}{3} \,0.1130 \,R_p^2 M \Omega_{\rm shedd},
\end{eqnarray}
we find
\begin{equation} \label{jdot1}
\frac{\dot J}{J} = \Big( \frac{3}{2} \Big)^3 \frac{1}{0.1130} \, 
	\frac{\dot M}{M}.
\end{equation}
Inserting this into~(\ref{rdot1}) yields
\begin{equation} \label{rdot2}
\frac{\dot R}{R} = 3  \frac{\dot M}{M} \,
\left( \Big( \frac{3}{2} \Big)^2 \frac{1}{0.1130} - 1 \right).
\end{equation}

Eqs.~(\ref{mdot3}), (\ref{jdot1}) and~(\ref{rdot2}) now completely 
determine the evolution of $M$, $J$ and $R$.  The equations can be 
simplified by defining the coefficients
\begin{eqnarray}
k_M & \equiv & 4 \pi \lambda_{\rm shedd} = 8.03 \nonumber \\
k_J & \equiv & \Big( \frac{3}{2} \Big)^3 \frac{1}{0.1130} = 29.86, \\
k_R & \equiv & 2 k_J - 3 = 56.73 \nonumber,
\end{eqnarray}
in terms of which eqs.~(\ref{mdot3}), (\ref{jdot1}) and~(\ref{rdot2})
can then be rewritten
\begin{eqnarray}
\frac{\dot M}{M} & = & - \frac{k_M}{\kappa} \frac{R}{M} \label{mdot4} \\[1mm]
\frac{\dot R}{R} & = & k_R \frac{\dot M}{M} \\[1mm]
\frac{\dot J}{J} & = & k_J \frac{\dot M}{M}.
\end{eqnarray}
The last two equations can be integrated immediately
\begin{eqnarray}
\frac{R}{R_{\rm crit}} & = & \left( \frac{M}{M_{\rm crit}} \right)^{k_R} 
	\label{rdot3} \\
\frac{J}{J_{\rm crit}} & = & \left( \frac{M}{M_{\rm crit}} \right)^{k_J}
	\label{jdot3}
\end{eqnarray}
Inserting~(\ref{rdot3}) into~(\ref{mdot4}) then yields
\begin{equation}
\dot M = - \frac{k_M R_{\rm crit}}{\kappa} 
\left( \frac{M}{M_{\rm crit}} \right)^{k_R}
\end{equation}
or
\begin{equation}
\int_{M_{\rm crit}}^M M^{-k_R} dM = - \frac{k_M}{\kappa}\, 
\frac{R_{\rm crit}}{M_{\rm crit}^{k_R}} \int_0^{t} dt,
\end{equation}
which can be integrated to
\begin{equation}
\frac{M}{M_{\rm crit}} = \left( 1 + \frac{(k_R - 1) k_M}{\kappa} 
	\Big(\frac{R}{M}\Big)_{\rm crit} t \right)^{1/(1-k_R)}, 
\end{equation}
where $-\infty \leq t \leq 0$ and $t=0$ corresponds to arrival at the
critical configuration.  Defining
\begin{equation} \label{tcrit}
t_{\rm crit} \equiv \frac{\kappa}{(k_R - 1) \,k_M} 
	\Big(\frac{M}{R}\Big)_{\rm crit},
\end{equation}
the last equation can be rewritten
\begin{equation} \label{mdot5}
\frac{M}{M_{\rm crit}} = \left(1 + \frac{t}{t_{\rm crit}} \right)^{1/(1-k_R)} 
\end{equation}
Note that $t_{\rm crit}$ is independent of the mass of the
star, and determines the evolutionary timescale of SMSs.  In cgs
units, it takes the value
\begin{equation} \label{tcrit_value}
t_{\rm crit} = \frac{c}{G} \,\frac{\kappa}{(k_R - 1) \,k_M} 
	\Big(\frac{GM}{c^2 R}\Big)_{\rm crit} = 8.8 \times 10^{11} \mbox{s}
\end{equation}
(compare Shapiro \& Teukolsky 1983, where a similar timescale has
been derived for SMSs that are stabilized by gas pressure rather than
rotation).  

\begin{figure*}
\epsscale{0.4} 
\plotone{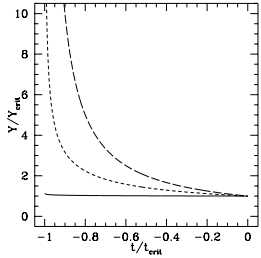}
\caption{Evolutionary tracks of the mass $M$ (solid line),
angular momentum $J/M^2$ (short dashed line) and radius $R/M$
(long dashed line), all normalized to their critical values at 
the onset of radial collapse (configuration $A$).  The critical time
$t_{\rm crit}$ is defined in eq.~(\protect\ref{tcrit}), and 
$t=0$ corresponds to the time at which the star reaches this onset
of instability.\label{fig5}}
\end{figure*}

Similar expressions for $R$ and $J$ can be found by
inserting~(\ref{mdot5}) into~(\ref{rdot3}) and~(\ref{jdot3}):
\begin{eqnarray}
\frac{R}{R_{\rm crit}} & = & 
	\left( 1 + \frac{t}{t_{\rm crit}} \right)^{k_R/(1-k_R)} 
	\label{rdot4} \\
\frac{J}{J_{\rm crit}} & = & 
	\left( 1 + \frac{t}{t_{\rm crit}} \right)^{k_J/(1-k_R)}
	\label{jdot4}
\end{eqnarray}
Using the identity $k_R = 2 k_J - 3$, we find that the dimensionless
quantities $J/M^2$ and $R/M$ evolve according to
\begin{eqnarray}
\frac{J/M^2}{(J/M^2)_{\rm crit}} & = & 
	\left( 1 + \frac{t}{t_{\rm crit}} \right)^{-1/2} \\
\frac{R/M}{(R/M)_{\rm crit}} & = & 
	\left( 1 + \frac{t}{t_{\rm crit}} \right)^{-1}
\end{eqnarray}
We plot $M$, $J/M^2$ and $R/M$ as a function of time in
Figure~\ref{fig5}.  Note that $J/M^2$ and $R/M$ evolve on the
timescale $t_{\rm crit}$, which is why we adopt this value for our
estimates in Section~\ref{Sec2}.  The mass $M$ and similarly $K$,
however, effectively evolve on a much longer timescale, which is a
consequence of the small numerical value of the exponent $1/(1-k_R) =
- 0.0179$.

\section{THE OUTCOME OF COLLAPSE: SPECULATIONS}
\label{coll}

Once the supermassive star becomes unstable, it will start to collapse
on a dynamical timescale.  The outcome of this collapse depends on
many factors, and can be determined only with a numerical,
three-dimensional hydrodynamics simulation in general relativity.
Such a calculation is only now underway (Baumgarte, Shapiro \& Shibata
1999).  The only dynamical, fully relativistic simulations of the
collapse of SMSs have been performed for nonrotating
configurations in spherical symmetry (see, for example, Shapiro \&
Teukolsky, 1979), for which black hole formation is well
established.  These calculations, however, do not shed much light on the
problem at hand, since rotation may play a crucial role in 
the dynamical evolution.  The issue of the final fate of a collapsing,
rapidly rotating SMS can therefore only be resolved by nonspherical,
relativistic numerical simulations.

In the meantime, however, we can attempt to assess crudely whether
this collapse can actually lead to the formation of a supermassive
black hole.  To do so, we assume that each mass shell conserves its
angular momentum during the collapse (cf.~the discussion at the end of
Section~\ref{Sec2}), and consider two criteria.

First, a particle can only be captured by a black hole if it is not be
repelled by the angular momentum barrier.  For simplicity, we take the
newly formed black hole to be a Schwarzschild black hole, in which
case a particle is in a capture orbit if
\begin{equation} \label{crit1}
l \leq 4 m,
\end{equation}
where $l$ is the particle's specific angular momentum, and $m$ the
mass of the black hole.  Here we ignore the effects of pressure,
assuming that in relativistic collapse the matter approaches the
nascent black hole supersonically by the time it enters the 
strong field domain.

Second, for any portion of the star to form a Kerr black
hole, the angular momentum $J$ of that portion cannot be larger than
the square of its mass $m$
\begin{equation} \label{crit2}
J/m^2 \leq 1.
\end{equation}
We crudely evaluate these two criteria neglecting post-Newtonian
corrections and nonspherical distortions due to rotation\footnote{ In
reality, interior distortions and angular momentum exchange may play
an important role late in the collapse, in particular if bars form, as
we argue below.}.

\begin{figure*}
\epsscale{0.4}
\plotone{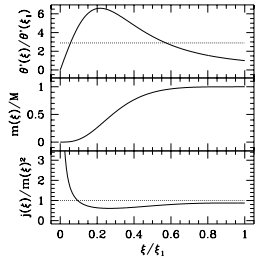}
\caption{Plots of the derivative of the Lane-Emden function $\theta(\xi)$,
the enclosed mass $m(\xi)$ and the enclosed angular momentum $J(\xi)$
divided by $m(\xi)^2$ as a function of $\xi$ for $n = 3$.  The
horizontal, dotted line in the top panel marks the threshold for
capture by a central black hole, and the dotted line in the bottom
panel marks the maximum angular momentum parameter for a Kerr black
hole.\label{fig6}}
\end{figure*}

The specific angular momentum of a particle at radius $r$ in the
equatorial plane, at the onset of instability, is
\begin{equation}
l = r^2 \Omega_{\rm shedd},
\end{equation}
where the critical orbital velocity $\Omega_{\rm crit}$ can be written
\begin{equation} 
\Omega_{\rm shedd}
	= \Big( \frac{2}{3} \Big)^{3/2} \Big( \frac{M}{R^3} \Big)^{1/2}
	=  \Big( \frac{2}{3} \Big)^{3/2} 
	\Big( \frac{R}{M} \Big)^{1/2}_{\rm crit}
	\frac{M}{R^2}.
\end{equation}
Here, $(R/M)_{\rm crit} \sim 450$ is the uniquely determined value at
the onset of instability (see Table~2).  Inserting the last two
equations into~(\ref{crit1}), we find
\begin{equation} \label{crit1_1}
\frac{m(r)}{r^2} \geq \frac{1}{4} \Big( \frac{2}{3} \Big)^{3/2}
		\Big( \frac{R}{M} \Big)^{1/2}_{\rm crit} \frac{M}{R^2},
\end{equation}
where we have assumed that the mass of the black hole, $m$, is the mass
enclosed by a sphere of radius $r$.

It is convenient to express the density and radius in terms of the 
dimensionless Lane-Emden functions $\theta$ and $\xi$
\begin{equation} \label{LE_var}
\begin{array}{lcr}
\rho & = & \rho_c \,\theta^n \\[1mm]
r    & = & a \,\xi,	
\end{array}
\end{equation}
where $a$ has units of length and is defined by
\begin{equation}
a = \left( \frac{(n+1) K \rho_c^{(1-n)/n}}{4 \pi} \right)^{1/2}
\end{equation}
(see, e.g., Shapiro \& Teukolsky 1983).  In terms of these quantities,
the mass $m(r)$ can be written
\begin{equation} \label{LE_mass}
m(r) = 4 \pi a^3 \rho_c \xi^2 |\theta'(\xi)|,
\end{equation}
where the prime denotes a derivative with respect to $\xi$.  For $n=3$,
the values of $\xi$ and $\theta'$ at the surface of the star are
$\xi_1 = 6.897$ and $|\theta'(\xi_1)| = 0.0424$.  Using~(\ref{LE_var})
and~(\ref{LE_mass}), we find that~(\ref{crit1_1}) reduces to a criterion
on $\theta'(\xi)$ alone:
\begin{equation} \label{crit1_2}
\frac{\theta'(\xi)}{\theta'(\xi_1)} \geq \frac{1}{4} 
		\Big( \frac{2}{3} \Big)^{3/2}
		\Big( \frac{R}{M} \Big)^{1/2}_{\rm crit} \sim 2.887,
\end{equation}
where we have adopted $(R/M)_{\rm crit} \sim 450$ for the critical
configuration.

We plot $\theta'(\xi)/\theta'(\xi_1)$ as a function of $\xi$ in the
top panel of Figure~6, and mark the threshold value $2.887$ by the
horizontal line.  A particle at a radius for which
$\theta'(\xi)/\theta'(\xi_1)$ is greater than this value can be
captured, if the regions interior to this radius have collapsed to a
black hole.  For $\xi > \xi_{\rm max} \sim 0.57 \,\xi_1$,
$\theta'(\xi)/\theta'(\xi_1)$ is less than the capture threshold, and
therefore the angular momentum barrier would prevent these regions from
being caught by a newly formed, interior black hole.  For the outermost
regions, this is not surprising, since the configuration is critically
rotating.  Therefore, the outer region may remain in orbit, perhaps in
a circumstellar disk, even if the rest of the star has
collapsed.  However, from the middle panel in
Figure~6 we find that this outer region only contains $~5\%$ of the
mass, while about 95\% of the mass could form a black hole.  This,
again, is a consequence of $n=3$ polytropes being extremely centrally
condensed.

According to Figure~6, the angular momentum barrier would also prevent
particles at very small radii, $\xi \lesssim 0.05 \xi_1$, from being
captured by a black hole interior to that radius.  However, for this
to be relevant, the initial black hole would have to be restricted to
a very small fractional size, and as we will see below, the second
criterion~(\ref{crit2}) does not allow such small black holes to form.

We may reverse the above argument and view eq.~(\ref{crit1_2}) as a
condition on $R/M$.  Since the left hand side,
$\theta'(\xi)/\theta'(\xi_1)$, has a maximum of about 6.6, $R/M$ has
to be smaller than about 2350 for black hole formation to be
possible. It is interesting that $R/M$ of the critical configuration
$A$ is only about five times smaller than this threshold compaction,
just barely allowing the supermassive star to form a supermassive
black hole. This argument suggests that primordial gas may have to
pass through a phase as a SMS where it can lose angular momentum
before it can possibly collapse to a SMBH.

We can similarly evaluate the second criterion, eq.~(\ref{crit2}), 
in terms of Lane-Emden functions.  The angular momentum $J(r)$ of 
the matter enclosed within radius $r$, rotating with the critical 
angular velocity, can be written
\begin{eqnarray}
J(r) & = & I(r) \, \Omega_{\rm shedd} \nonumber \\[1mm] 
	& = & \big( \frac{2}{3} \big)^{3/2} 
	\big( \frac{R}{M} \big)_{\rm crit}^{1/2} \frac{M}{R^2}
	\frac{8 \pi}{3} \int_0^r \rho r^4 d r \\[1mm]
	& = &  \big( \frac{2}{3} \big)^{5/2} 
	\big( \frac{R}{M} \big)^{1/2}_{\rm crit}
	(4 \pi)^2 a^6 \rho_c^2 |\theta'(\xi_1)| 
	\int_0^{\xi} \theta^n \xi^4 d\xi. \nonumber
\end{eqnarray}
Dividing this by the square of the mass, eq.~(\ref{LE_mass}), then
yields the condition
\begin{equation} \label{crit2_1}
1 \geq \frac{J(\xi)}{m(\xi)^2} = 
	\big( \frac{2}{3} \big)^{5/2} \big( \frac{R}{M} \big)^{1/2}_{\rm crit}
	\frac{|\theta'(\xi_1)|}{\xi^4 |\theta'(\xi)|^2} 
	\int_0^{\xi} \theta^n \xi^4 d\xi.
\end{equation}
In the bottom panel of Figure~6 we show the right hand side of this
equation for a $n=3$ polytrope and for $(R/M)_{\rm crit} = 450$.  The
dotted, horizontal line is the critical value of unity.  A region
inside a radius $\xi$ can form a Kerr black hole only if
$J(\xi)/m(\xi)^2$ is less than unity.  This condition is satisfied
everywhere except for the innermost regions $\xi < \xi_{\rm min} \sim
0.1 \,\xi_1$.  At the surface, $\xi = \xi_1$, we recover the value
$(J/M)^2_{\rm crit} = 0.876$ found in eq.~(\ref{crit_values}).

Note that for any polytrope, $m$ is proportional to $r^3$ for small radii
$r$ and $J$ is proportional to $m(r) \Omega r^2 \sim
\Omega r^5$.  Therefore, $J/m^2$
scales like $\Omega/r$ close to the center, which prevents the
formation of arbitrarily small black holes from rotating polytropes.
In our case, the region inside $\xi_{\rm min} \sim 0.1\,
\xi_1$, containing about 4\% of the total mass, defines the minimum
mass which can collapse to form an initial black hole.  Since
$\theta'(\xi)/\theta'(\xi_1)$ exceeds the threshold for $\xi \gtrsim
\xi_{\rm min}$, such a ``minimal'' seed black hole could accrete
further material from the star.

Finally, we note that a bar instability may form during the
collapse of the star.  The criterion for the formation of a bar
on a dynamical timescale is
\begin{equation}
\left( \frac{T}{|W|} \right)_{\rm bar} \gtrsim 0.27
\end{equation}
(see, e.g., Chandrasekhar 1969; Shapiro \& Teukolsky 1983).  Since $M$
and $J$ are approximately conserved during the collapse, $T/|W|$
scales with $R^{-1}$, so a bar will start to form when
\begin{equation}
\frac{R_{\rm bar}}{R_{\rm crit}} = \left( \frac{T}{|W|} \right)_{\rm crit}
\left( \frac{T}{|W|} \right)_{\rm bar}^{-1} \sim \frac{0.009}{0.27} 
= \frac{1}{30}.
\end{equation}
The value of $R/M$ at bar formation is then
\begin{equation} \label{R_bar}
\left( \frac{R}{M} \right)_{\rm bar} \sim 
\frac{1}{30} \left( \frac{R}{M} \right)_{\rm crit} \sim 15.
\end{equation}
Eg.~(\ref{R_bar}) suggests that the collapsing star may form a
nonaxisymmetric bar before it forms a black hole.  This is an
important result, since such a bar may result in a quasiperiodic
emission of gravitational waves (cf.~Smith, Houser \& Centrella 1996).
The frequency of these waves can be estimated from the expected bar
rotation rate
\begin{eqnarray}
f_{\rm bar} & \sim & \frac{\Omega_{\rm bar}}{2\pi} 
	\sim \frac{1}{2 \pi} \left( \frac{M}{R_{\rm bar}^3} \right)^{1/2} 
	= \frac{1}{2 \pi}  \left( \frac{M}{R_{\rm bar}} \right)^{3/2}
	\frac{1}{M} \nonumber \\
	& \sim & 5 \times 10^2 \,\mbox{Hz} \,
	\left( \frac{M}{M_{\odot}} \right)^{-1}.
\end{eqnarray}
For a SMS of $10^6 M_{\odot}$ this yields a frequency of $5 \times
10^{-4}$ Hz, which is in the range in which LISA is expected to be
most sensitive (see, e.g., Thorne 1995).  The frequency increases at
later stages of the collapse, when $M/R_{\rm bar}$ becomes larger.  The
strength of the gravitational wave signal can be crudely estimated to
be
\begin{eqnarray}
h & \sim & \frac{\ddot Q}{d} \sim \frac{M R_{\rm bar}^2 f_{\rm bar}^2}{d}
	\sim \frac{M}{d} \, \frac{1}{4 \pi^2} \, \frac{M}{R_{\rm bar}}  
	\nonumber \\
	& \sim & 10^{-25} \left(\frac{M}{M_{\odot}} \right)
	\left( \frac{d}{1 \mbox{Gpc}} \right)^{-1},
\end{eqnarray}
where $Q$ is the star's quadrupole moment, and $d$ is the distance,
which we scale to 1 Gpc (the Hubble distance is $\sim 3$ Gpc).  The
signal strength increases with $M/R_{\rm bar}$ at late stages of the
collapse.  Apart from any bar, the implosion itself will be
nonspherical, due to rotation, and will result in a gravitational wave
burst. The collapse of a supermassive star may therefore be a very
promising candidate for detection by space based gravitational wave
detectors (LISA Pre-Phase A report 1995).

Obviously, the arguments presented in this section are very crude and
do not replace a fully self-consistent relativistic hydrodynamical
calculation.  Our arguments nevertheless suggest that, upon
reaching the onset of instability, supermassive stars may form a
supermassive black hole, containing a large part of the mass, and
leaving only a few percent of the initial mass outside of the black
hole, most likely in the form of a disk.  We furthermore anticipate that
a bar may form during the collapse, which may lead
to the emission of quasiperiodic gravitational waves.

\section{SUMMARY}
\label{Summary}

We have launched a fully relativistic study of the formation of SMBHs
via the collapse of SMSs.  In this paper we study the quasistationary,
secular evolution of SMSs up to the critical configuration at which
radial instability sets in and focus on the effects of rotation and
general relativity.  We identify the critical configuration and its
characteristic parameters $R/M$, $T/|W|$ and $J/M^2$.  These
parameters are independent of the mass of the SMS, and are therefore
universal constants.  The subsequent implosion, starting from this
universal critical configuration, is therefore also uniquely
determined and should produce a unique gravitational waveform.  We
compare results from an analytic, approximate treatment and a fully
relativistic, numerical calculation, and find good agreement.  We
furthermore solve analytically for the time evolution of these
parameters up to the onset of instability.

Identifying the critical configuration at the onset of instability is
interesting for its own sake.  More importantly, however, this
configuration will be adopted as initial data for future numerical
simulations of SMS collapse (Baumgarte, Shapiro \& Shibata 1999).  In
this paper we assemble qualitative arguments to anticipate the outcome of
the collapse and find that the formation of a SMBH containing an
appreciable fraction of the mass is not ruled out.  Our arguments suggest
that a transient phase as a SMS may be an efficient way for primordial gas
to lose sufficient angular momentum in order to overcome the Kerr angular
momentum barrier to forming a SMBH.

Loeb \& Rasio (1994) have emphasized that SMBHs appear to have a
minimum ``observed'' mass of $\sim 10^6 M_{\odot}$.  Our crude
arguments may help to explain this minimum in the context of the SMS
formation scenario.  SMSs with masses less than this value will not
evolve in a quasistationary manner up to our critical configuration
$A$, because some of our assumptions break down.  For example, nuclear
burning may cause the star to explode (Appenzeller \& Fricke 1972,
Fuller, Woosley \& Weaver, 1986).
Electron-positron pairs may also be important for small masses, and
may destabilize the star (see Zel'dovich \& No\-vi\-kov 1971) before it
reaches a small enough value of $R/M$ to overcome the angular momentum
barrier.  On the other hand, all SMSs with $M \gtrsim 10^6 M_{\odot}$
will evolve in a quasistationary manner until reaching the critical
configuration $A$.  The subsequent collapse will likely give rise to a
SMBH of almost the same mass as the progenitor SMS.

\acknowledgments

This paper was supported in part by NSF Grants AST 96-18524 and PHY
99-02833 and NASA Grant NAG5-7152 to the University of Illinois at
Urbana-Champaign.

\end{document}